\documentclass[10pt,twocolumn]{article} 
\usepackage{wscad2009}
\usepackage[dvips]{graphicx}
\usepackage{url}

\begin{document}
\title{Couillard: Parallel Programming via Coarse-Grained Data-Flow Compilation}

\author{Leandro A. J. Marzulo, Tiago A. O. Alves, Felipe M. G. França\\
Universidade Federal do Rio de Janeiro\\ Programa de Engenharia de Sistemas e
Computação, COPPE \\ Rio de Janeiro, RJ, Brasil \\ \{tiagoaoa, lmarzulo,
felipe\}@cos.ufrj.br\\
% Para autores que pertencem a uma mesma instituição, 
% omita as seguintes linhas até o último ``}''.
% Neste caso use $^{1}$ para colocar um sobrescrito nos autores e
% respectivos endereços de e-mail
% Para adicionar autores com endereços diferentes coloque um ``\and'', 
% como no caso do segundo autor a seguir:
\and
Vítor Santos Costa\\
Universidade do Porto\\
Departamento de Ciência de Computadores\\ Porto, Portugal\\ 
vsc@dcc.fc.up.pt\\
}

\maketitle
\thispagestyle{empty}

\begin{abstract}
Data-flow is a natural approach to parallelism. However, describing dependencies
and control between fine-grained data-flow tasks can be complex and present unwanted
overheads. TALM (TALM is an Architecture and Language for Multi-threading) 
introduces a user-defined coarse-grained parallel data-flow
model, where programmers identify code blocks, called super-instructions, to be
run in parallel and connect them in a data-flow graph. TALM has been implemented
as a hybrid Von Neumann/data-flow execution system: the \emph{Trebuchet}. We have
observed that TALM's usefulness largely depends on how programmers specify and
connect super-instructions. Thus, we present \emph{Couillard}, a full compiler that
creates, based on an annotated C-program, a data-flow graph and C-code
corresponding to each super-instruction. We show that our toolchain allows one to benefit 
from data-flow execution and explore sophisticated parallel programming techniques, with small effort. To evaluate
our system we have executed a set of real applications 
on a large multi-core machine. Comparison with popular parallel programming methods
shows competitive speedups, while providing an easier parallel programing
approach.
\end{abstract}

\Section{Introduction}

%Data-flow programming provides a natural approach to parallelism,
%where instructions execute as soon as their input operands are
%available~\cite{2468,78583,10.1109/SBAC-PAD.2008.29,Swanson2003}. 
%However, fully describing the data-flow graph of an application can
%be an arduous task, since it involves specifying all control (loops
%and branches, for example) using data-flow instructions that will 
%send operands to the correct graph path. In fact, this is one of the main enablers of
%parallelism exploitation in data-flow: control is not changed just by altering
%a program counter. In a loop, for example, branches are taken individually by 
%each piece of data that is needed inside the loop body, allowing decoupled execution
%of independent portions of the loop.. Actually, in dynamic data-flow,
%we may even have independent instructions 
%from multiple iterations running simultaneously.
%Nevertheless, although data-flow is
%nowadays a fundamental technique in computer design~\cite{tomasulo}, it has
%not so far taken a mainstream role in parallel programming.

Data-flow programming provides a natural approach to parallelism, where instructions execute as soon as their input operands are available~\cite{2468,78583,10.1109/SBAC-PAD.2008.29,Swanson2003}. Actually in dynamic data-flow, we may even have independent instructions from multiple iterations on a loop running simultaneously, as parts of the loop may run fast than others and reach next iterations.  Therefore it is complex to describe control in data-flow, since instructions must only proceed to execution when operands from the same iteration match. However, this difficulty is compensated by the amount of parallelism exploited this way

TALM (TALM is an Architecture and Language for Multi-threading)
\cite{trebwammca,trebuchet_ijhpsa,wscad09trebuchet} is an execution 
model designed to exploit the advantages of data-flow in multi-thread 
programming. A program in TALM is comprised of code blocks called 
\emph{super-instructions} and simple instructions connected in a graph 
according to their data dependencies (i.e. a data-flow graph). To 
parallelize a program using the TALM model, the programmer  
marks portions of code that are to become super-instructions and describe 
their dependencies. With this approach, parallelism comes naturally from data-flow execution.

The major advantage of TALM is that it provides a coarse-grained
parallel model that can take advantage of data-flow. It is also a very
flexible model, as the main data-flow instructions are
available, thus allowing full compilation of control in a data-flow fashion.
This gives the programmer the latitude to choose from
coarser to more fine-grained execution strategies. This
approach contrasts with previous work in data-flow programming~\cite{2468,78583,Swanson2003},
which often aimed at hiding data-flow execution from the programmer. 
%Those systems were,
%however, full data-flow architectures that did not become popular.

A first implementation of TALM, the \emph{Trebuchet} system, has been
developed as a hybrid Von Neumann/data-flow execution system for
thread-based architectures in shared memory platforms. \emph{Trebuchet}
emulates a data-flow machine that supports both simple instructions and super-instructions. 
Super-instructions are compiled as separate functions that are called by
the runtime environment, while regular instructions are interpreted upon execution. 
Although \emph{Trebuchet} needs to emulate data-flow
instructions, experience showed most running time is within our super-instructions. 
Initial results show the parallel engine to be competitive with
state-of-the-art parallel applications using OpenMP, both in terms of
base performance, and in terms of speedups~\cite{trebwammca,trebuchet_ijhpsa,wscad09trebuchet}. 
On the other hand,
parallelism for for simple SPMD (Single-Program Multiple Data) 
applications can be explored quite well with
tools such as OpenMP. The main benefits exploited by TALM become apparent when
experimenting with applications that require more complex techniques,
such as software pipelining or speculative execution.

The usefulness of TALM clearly depends on how the programmer can
specify and connect super-instructions together, 
including the complex task of describing control using data-flow instructions.
We therefore introduce \emph{Couillard}, a \texttt{C}-compiler designed to
compile TALM annotated \texttt{C}-programs into a data-flow
graph, including the description of program control using dynamic data-flow.
\emph{Couillard} is designed to insulate the programmer from the details of 
data-flow programming. By requiring the programmer to just annotate the code with the
super-instruction definitions and their dependencies, \emph{Couillard} greatly
simplifies the task of parallelizing applications with TALM.

This work makes two contributions:
\begin{itemize}
\item We define the TALM language, as an extension of ANSI C and present a full 
implementation of the \emph{Couillard} Compiler, which generates data-flow graphs
and super-instruction code for TALM. 
\item We evaluate the performance of \emph{Couillard} on two state-of-the-art
  PARSEC~\cite{bienia11benchmarking} benchmarks. We demonstrate that \emph{Trebuchet} and
  \emph{Couillard} allows one to explore complex parallel programing techniques,
  such as non-linear software pipelines and hiding I/O latency. Comparison with popular parallel
  programming models, such as Pthreads~\cite{pthreads}, OpenMP~\cite{615542} and
  Intel Thread Building Blocks~\cite{tbb} shows that our approach is not
  just competitive with state-of-the-art technology, but that in fact
  can achieve better speedups by allowing one to easily exploit a
  sophisticated design space for parallel programs.
\end{itemize}

The paper is organized as follows. In Sect.~\ref{talm} we briefly
review TALM architecture and its implementation, the \emph{Trebuchet}. 
In Sect.~\ref{compilation} we describe TALM language and Couillard implementation. In
Sect.~\ref{results} we present performance results on the two
PARSEC benchmarks. In Sect.~\ref{related} we discuss some related works. 
Last, we present our conclusions and discuss
future work.

\Section{TALM and Trebuchet}\label{talm}

%Most modern computers provide an Instruction Set Architecture (ISA)
%where programs execution is guided by control-flow, i.e., by following
%the Von Neumann model.  In fact, most computers internally implement
%out-of-order execution mechanisms, such as the Tomasulo algorithm
%\cite{tomasulo}. These mechanisms often rely on data-flow to extract
%instruction level parallelism (ILP). However, in modern multi and
%many-core processors, exploring ILP is not sufficient, and exploring
%thread-level parallelism (TLP) has become a major requirement if one
%wants to take full advantage of modern architectures.

TALM \cite{trebwammca,trebuchet_ijhpsa,wscad09trebuchet} allows application
developers to take advantage of the possibilities available in the
data-flow model in current Von Neumann architectures, in order to
explore TLP in a more flexible way.  TALM ISA sees applications in
the form of a data-flow graph that can be run in parallel.   

A main contribution of TALM is that it enables programmers to
introduce user-defined instructions, the so called {\it super-instructions}. 
TALM assumes a contract with the programmer
whether she or he guarantees that execution of the super-instruction
can start if all inputs are available, and where she or he guarantees
to make output arguments available as soon as possible, but not
sooner. Otherwise, TALM has no information on the semantics of
individual super-instructions, and indeed imposes no
restrictions. Thus, a programmer can use shared memory in
super-instructions without having to inform TALM. Although this
requires extra care from the programmer, the advantage is that TALM
allows easy porting of imperative programs and easily allows program
refinement.

TALM has been implemented for multi-cores as a hybrid Von Neumann/data-flow 
execution system: the \emph{Trebuchet}. \emph{Trebuchet} is in fact a data-flow virtual machine that has a set of
data-flow processing elements (PEs) connected in a virtual network. Each PE is associated with a
thread at the host (Von Neumann) machine. When a program is executed on \emph{Trebuchet},
instructions are loaded into the individual PEs and fired according to
the Data-flow model. Independent instructions will run in parallel if
they are mapped to different PEs and there are available cores at the
host machine to run those PEs' threads simultaneously.

\emph{Trebuchet} is a Posix-threads based implementation of
TALM. It loads super-instructions as a dynamically linked library. At
run-time, execution of super-instructions is fired by the virtual
machine, according to the data-flow model, but their interpretation
comes down to a procedure call resulting in the direct execution of
the related block.

\begin{figure}[h!ptb]
\begin{center}
\includegraphics[width=\linewidth]{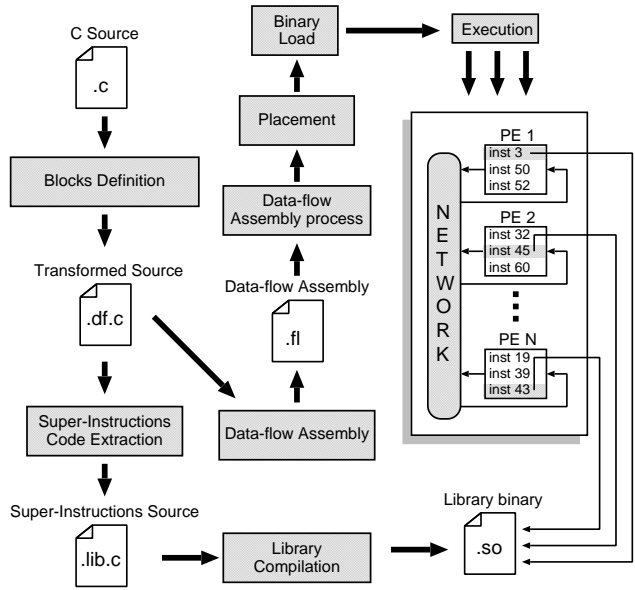}  
\caption{\label{workflow}Work-flow to follow when writing parallel applications
with \emph{Trebuchet}.}
\end{center}
\end{figure}

\emph{Trebuchet} may either rely solely on static scheduling of instructions 
among PEs or may also use work-stealing as a tool against imbalance. 
The work-stealing algorithm employed by \emph{Trebuchet} is based on the 
ABP algorithm \cite{Arora2001a}, the main difference being that the algorithm 
developed for \emph{Trebuchet} provides a FIFO double-ended queue (deque) 
instead of a LIFO one, as is the case for the ABP algorithm. The FIFO order is chosen
so that older instructions have execution priority, which is desirable for the applications
we target at this moment.

Figure \ref{workflow} shows the work-flow to be followed in order to
parallelize a sequential program and execute on \emph{Trebuchet}.
Initially, blocks that will form super-instructions are defined. Then,
a super-instruction code extraction is performed to transform all blocks
into functions that will collect input operands from \emph{Trebuchet}, process and
return output operands.
Profiling tools may be used in helping to determine which portions of
code are interesting candidates for parallelization.

In the next step, the transformed blocks are compiled into a dynamic
library, which will be available to the abstract machine
interpreter. Then, a data-flow graph connecting all blocks is defined
and the data-flow assembly code is generated. The code may have both
super-instructions and simple (fine-grained) instructions. TALM
provides all the standard data and control instructions that one would
expect in a dynamic data-flow machine.

Last, a data-flow binary is generated from the assembly, processor
placement is defined, and the binary code is loaded and executed. As
said above, execution of simple instructions requires full
interpretation, whereas super-instructions are directly executed on
the host machine.

In \cite{trebuchet_ijhpsa,wscad09trebuchet} TALM was used to
parallelize a set of 7 applications: a matrix determinant calculation,
a matrix multiplication application, a ray tracing application, Equake
from SpecOMP 2001, IS from NPB3.0-OMP, and also LU and Mandelbrot from
the OpenMP Source code Repository \cite{10.1109/EMPDP.2005.41}. The
achieved speedups for 8 threads, in relation to the sequential
versions were, respectively 2.52, 4.16, 4.39, 3.61, 3.00, 2.19 and
7.16. On the other hand, OpenMP versions of those benchmarks have 
provided speedups of 1.96, 4.15, 4.39, 3.40, 3.11, 2.19 and 7.13. 
These results are very promising, and show
that \emph{Trebuchet} can be very competitive with OpenMP for regular
applications.

\emph{Trebuchet} provides a natural platform for experimenting with
advanced parallel programming techniques. In \cite{trebwammca} 
a thread-level speculation model based on
optimistic transactions with ordered commits was created for TALM
and implemented in \emph{Trebuchet}. Execution of speculative instructions is done
within transactions, each one formed by one speculative instruction
and its related \texttt{Commit} instruction. 
Transactions will have access only to local copies of the
used resources. Once they finish running, if no conflicts are found,
local changes will be persisted to global state by \texttt{Commit}
instructions, associated with each speculative instruction. In case
conflicts are found in a speculative instruction $I$, local changes
will be discarded and $I$ will have to be re-executed.

Using speculative execution liberates the programmer to
consider only explicit dependencies while guaranteeing correct execution 
of coarse-grained tasks. Moreover, the speculation mechanism does not
demand centralized control, which is a key feature for upcoming many-core
systems, where scalability has become an important concern.
To evaluate the speculation system, a bank
server simulator artificial application was implemented to simulate
scenarios varying computation load, transaction size, speculation
depth, and contention. Results of execution of this application with
up to 24 threads in a 24-core machine suggest that there is a wide
range of situations where speculation can be very effective and indeed
achieve speedups close to the ideal case.

\Section{Compilation}
\label{compilation}

The data-flow model exposes thread-level parallelism by taking
advantage of how data is exchanged between processing elements. In
this vein, programming in TALM is about identifying
parallel tasks and how data is consumed and produced between them. The
initial \emph{Trebuchet} implementation provided an execution environment for
multi-cores, plus an assembler and loader.  It was up to the
programmer to code super-instructions in the library and to write
TALM assembly code linking the different instructions together
and specifying control trough data-flow instructions,
not always a trivial task.

In this work we propose \emph{Couillard}, a \texttt{C}-compiler
for data-flow style execution. With \emph{Couillard}, the programmer annotates
blocks of code that are going to become super-instructions, and
further annotates the program variables that correspond to their
inputs and outputs. \emph{Couillard} then produces the \texttt{C}-code
corresponding to each super-instruction to be next compiled as a shared 
object to the target architecture and loaded by \emph{Trebuchet}. Moreover, \emph{Couillard}
generates TALM assembly code to connect all super-instructions according to the
user's specification. This assembly code represents the actual data-flow graph 
of the program. Moreover, control
constructs such as loops and if-then-else statements that are not within 
super-instruction will also be compiled to TALM assembly code. This
assembly code will then be used by \emph{Trebuchet} to guide execution,
following the data-flow rules.

\emph{Couillard} front-end uses PLY (Python Lex-Yacc) \cite{beazley} and a grammar that is a subset of ANSI-C 
extended with super-instruction constructs. \emph{Couillard} back-end, to generates 
TALM assembly code for TALM, super-instructions C-code (to be compiled 
into a dynamically linked library) and a graph representation of the program,
using Graphviz notation \cite{graphviz}.

\SubSection{Front-end}

We assume that super-instructions take most of the running time of an
application, as regular instructions are mostly used to describe the
data and control relations between super-instructions. Since super-instruction
code will be compiled using a regular \texttt{C}-compiler and regular
instructions tend to be simple, \emph{Couillard} does not need to support the
full ANSI-\texttt{C} grammar. 
%However, \emph{Couillard} was designed to be easily extended, if
%necessary.

\emph{Couillard}, therefore adopts a subset of the ANSI-C grammar extended to support data-flow
directives relative to super-instructions and their dependencies. We have also changed 
the syntax of variable declaration and access, which is necessary to parallelize super-instructions. 
The compiler front-end produces an AST (Auxiliary Syntax Tree) that will be processed
to generate a data-flow graph representation.

\subsubsection{Blocks and Super-Instructions}\label{block}

The annotation pair \texttt{\#BEGINBLOCK} and \texttt{\#ENDBLOCK} is
used to mark blocks of code that will \emph{not} be compiled to
data-flow. Those blocks usually contain include files, auxiliary function
definitions, and global variables declarations, to be used by
super-instruction code in the dynamic library.

Super-instruction annotation is performed according to the following
syntax:

\begin{verbatim}
treb_super <single|parallel> input(<inputs_list>)
                              output(<output_list>)
#BEGINSUPER
    ...
#ENDSUPER
\end{verbatim}

Super-instructions declared as \texttt{single} will always have only one 
instance in the data-flow graph, while
instructions declared as \texttt{parallel} may have multiple
instances that can run in parallel, depending on the placement and
availability of resources at the host machine. In the example of Fig. \ref{pipeline}
(described in more details in Section \ref{examples}), we have
single super-instructions at the beginning and end of the
computation. In contrast, the inner code corresponds to parallel
super-instructions. 

\subsubsection{Variables}

\emph{Couillard} requires the programmer to specify how variables connect the
different super-instructions. More precisely, all variables used as
inputs or outputs of super-instructions must be previously declared
to guarantee that data will be exchanged correctly between
instructions (without loss due to wrong type castings). Also, output
variables used on parallel super-instructions must be declared as follows:

\begin{verbatim}
treb_parout <type> <identifier>;
\end{verbatim}

The \emph{Storage Classifier} \texttt{treb\_parout} is used because parallel super-instructions, 
in general, have multiple instances, Therefore, output variables of parallel super-instructions will also
have multiple instance, one for each instance of the parallel super-instruction. 

\begin{figure}[thpb]
\begin{center}
\includegraphics[width=\linewidth]{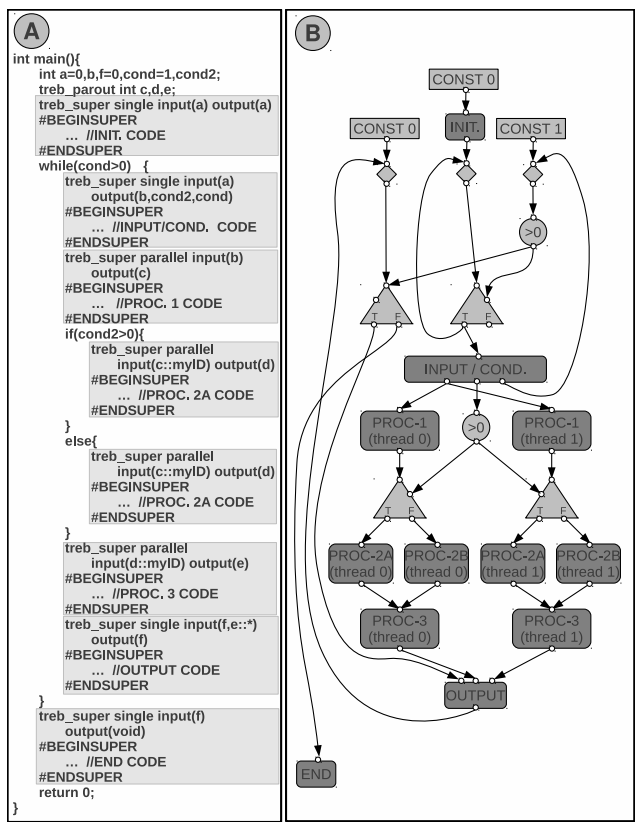}  
\caption{\label{pipeline}Example of non-linear parallel pipeline with TALM.}
\end{center}
\end{figure}

When using a
\texttt{treb\_parout} variable as input to another super-instruction (or
even in external \texttt{C}-code) it is necessary to specify the
instance that is being referenced. To do so, \emph{Couillard} provides the following
syntax:

\begin{verbatim}
<identifier>::< NUMBER |
                * | 
                mytid | 
                (mytid + NUMBER) | 
                (mytid - NUMBER) |
                lattid>
\end{verbatim}

Consider a variable named $x$. The notation $x::0$ refers to instance
$0$ of variable $x$, while $x::*$ refers to all instances of this
variable (this provides an useful abstraction when a super-instruction
can receive input from a number of sources). Also, it is often
convenient to refer to the instance for the current (parallel)
super-instruction. If $x$ is used as input to another parallel
super-instruction, we can select $x$ through the expression $x::mytid$.
To illustrate this situation, in the example of Fig. \ref{pipeline}, 
each instance $k$ ($0 \leq k \leq 1$, since there are 2 instances of each parallel super-instruction) 
of \texttt{Proc-2A} receives as input $c::k$, produced by \texttt{Proc-1}. Expression with $+$ and $-$ 
are also allowed with $mytid$. For example, if a parallel super-instruction $X$
produces operand $a$ and another parallel super-instructions $Y$ uses specifies
$a::(mytid-1)$ as input, it means that for a task $i$, $Y.i$ will receive $a$ from $X.(i-1)$.
Last, the reserved word $lasttid$ refers to the last instance of a parallel super-instruction and can be used to specify
inputs to parallel and single super-instruction.

For the cases were there are dependencies between instances of the same parallel super-instructions we can specify
input variables using the following construct:

\begin{verbatim}
local.<identifier>::<(mytid + NUMBER) | 
                     (mytid - NUMBER)>
\end{verbatim}

For example, if we state that a parallel super-instruction $s$ produces operand $o$ and receives $local.o::(mytid-2)$, 
it means that $s.i$ (instance $i$ of $s$) depends on $s.(i-2)$. Moreover, it means that $s.0$ and $s.1$ do not have local dependencies.
We can also specify operands that will be sent only to those independent instances of $s$. We use the following syntax:

\begin{verbatim}
starter.<identifier>::< NUMBER |
                * | 
                mytid | 
                (mytid + NUMBER) | 
                (mytid - NUMBER) |
                lattid>
\end{verbatim}

In the former example if we also define $starter.c$ as an input of $s$, only $s.0$ and $s.1$ will receive this operand. A practical
example of use of this constructs is to serialize distributed I/O operation to hide I/O latency, explained in Section \ref{examples}.

The rationale to describe parallel code in super-instructions is
simple. The developer first divides the code in blocks that can be run
in parallel. Initialization and termination blocks will most often be
\texttt{single}, whereas most of the parallel work will be in
\texttt{parallel} blocks. The programmer next specifies how the blocks
communicate. If the communication is purely control-based the programmer
should further add an extra variable to specify this connection (a
common technique in parallel programming). Note that the programmer
still has to prevent data races between blocks unless speculative 
execution is used (which is not yet supported by the compiler).

\SubSection{Back-end}

After generating an Abstract Syntax Tree (AST) of a program, \emph{Couillard} 
produces its corresponding data-flow graph. From this graph, it generates
three output files:

\begin{enumerate}
 \item A \texttt{.dot} file describing the graph in the Graphviz \cite{graphviz} notation. 
This file will be used to create an image of that graph, using 
the Graphviz toolchain. Although a Graphviz graph is not needed by \emph{Trebuchet}, it may be useful for academic 
purposes or to provide a more intelligible look of the produced graph to the programmers that want to 
and perform manual adjustments to their applications. 

 \item A \texttt{.fl} file describing the graph using TALM's ISA. This file will be the input to \emph{Trebuchet}'s
Assembler, producing the \texttt{.flb} binary file that will be loaded into \emph{Trebuchet}'s Virtual Machine.

 \item A \texttt{.lib.c} file describing the super-instructions as functions, in C-code, to be compiled 
as a dynamically linked library, using any regular C-compiler. All inputs and outputs variables described
with \emph{Couillard} syntax are automatically declared and initialized within the generated function.
Notice also that the super-instruction body does not need to
parsed by \emph{Couillard}. It is just treated as the value of a super-instruction node at the AST representation. This
allowed us to focus only on the instructions necessary to connect super-instructions in a coarse-grained 
data-flow graph.

\end{enumerate}

\SubSection{Auxiliary Functions and Command Line Arguments}

The functions, \texttt{treb\_get\_tid()} and
\texttt{treb\_get\_n\_tasks()}, have been added to \emph{Trebuchet}
virtual machine and they can be called inside super-instructions
code. The former returns the \emph{thread id} of that
super-instruction's instance, while the later returns the \emph{number
  of threads}. Those functions can be used to identify the portion of
work to be done by each instance.

\begin{figure}[h!ptb]
\begin{center}
\includegraphics[width=\linewidth]{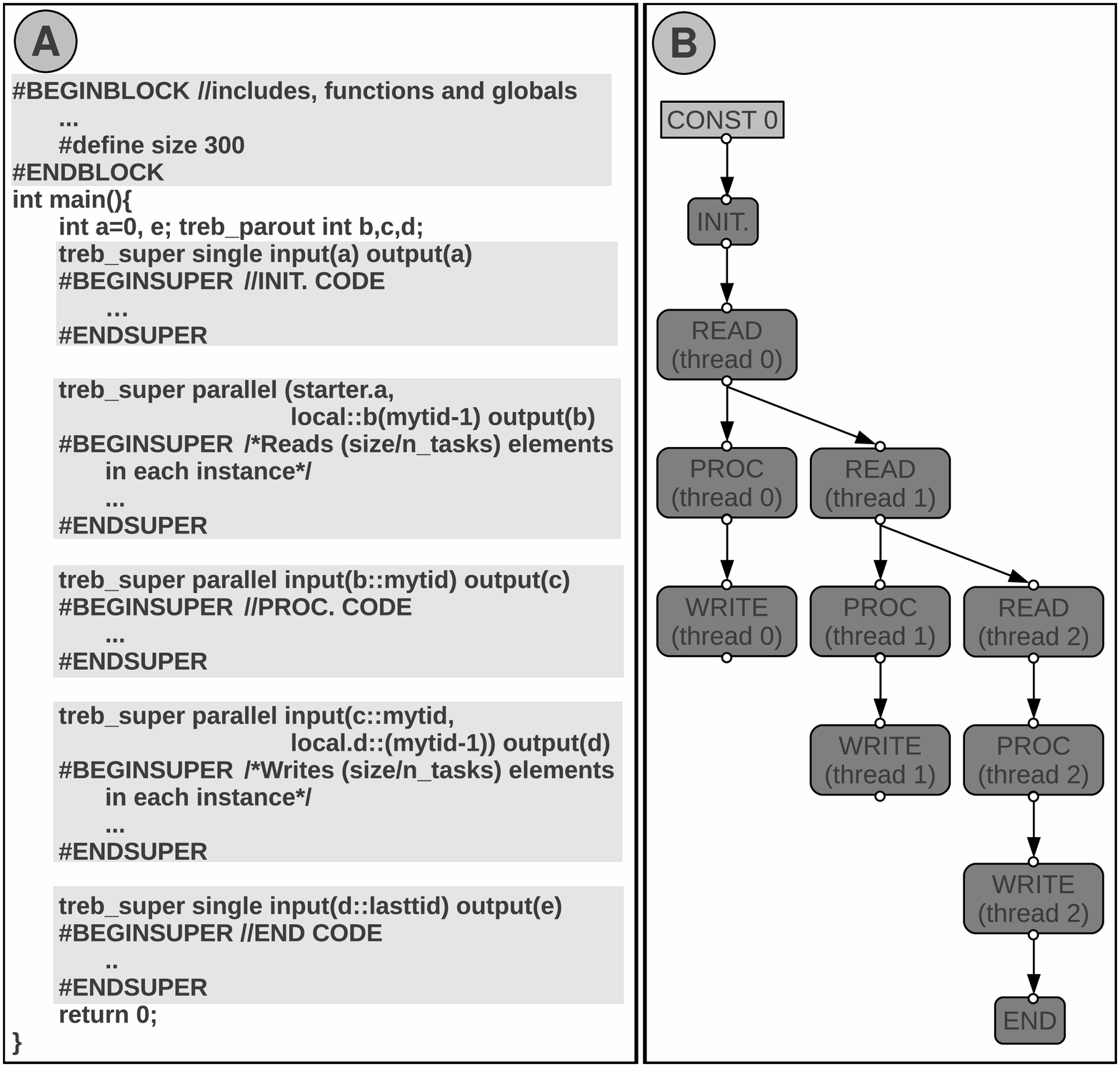}  
\caption{\label{hideio}Example of how to hide I/O latency with TALM.}
\end{center}
\end{figure}

In our system, applications are executed within the \emph{Trebuchet} virtual machine. 
Therefore, command line argument variables cannot be declared within the application's code.
They need to be passed trough \emph{Trebuchet}'s command line. Thus, \emph{Trebuchet} stores a vector of command line arguments
and the number of arguments at \texttt{treb\_superargv} and \texttt{treb\_superargc} variables, respectively. 
Then, \emph{Couillard} declares those variables as \texttt{extern} when generating the \texttt{.lib.c} file, 
meaning that programmers can access those arguments within super-instructions' body.

\SubSection{Illustrative Examples} \label{examples}

Figure \ref{hideio} provides an example of how TALM high-level language is used to hide I/O latency in a parallel application.
In this example we assume that 300 elements need to be read from a file, processed and then the result must be
written in an output file. In pane $A$ we can see the different steps to be performed by super-instructions 
(inner code not shown): \emph{(i)} initialization of variables and FILE pointers, \emph{(ii)} reading,
\emph{(iii)} processing, \emph{(iv)} writing and \emph{(v)} closing of files. Pane $B$ shows the associated data-flow
graph, generated by \emph{Couillard}.

One can notice that reading and writing
stages are described as parallel super-instructions, but since there are local inputs, they will be executed serially 
(although spread among different PEs). This construct allows the execution of each processing task to start as soon as the corresponding 
read operation has finished, instead of waiting for the hole read. It also allows writing the results of each processing
task $i$ without having to wait for tasks $x$, where $x<i$, to finish.

Figure \ref{pipeline} provides an example of how to use TALM high-level language to
describe a non-linear parallel pipeline. The example is a skeleton code
of an application that reads a file containing a bag of tasks to be
processed and writes the results to another file. The processing phase
can be divided in 3 stages (\texttt{Proc-1}, \texttt{Proc-2} and
\texttt{Proc-3}). The processing task, \texttt{Proc-2}, was divided in
two different tasks (\texttt{Proc-2A} and \texttt{Proc-2B}), that are
executed conditionally. Figure \ref{pipeline} (pane $A$) shows TALM
annotations, while the corresponding data-flow graph for 2
threads, generated by the \emph{Couillard} compiler, is shown in
Fig. \ref{pipeline} (pane $B$).

\Section{Experiments and Results}
\label{results}

Our goal is to obtain good performance in real applications and evaluate the
TALM for complex parallel programming. We study how our model
performs on two state-of-the-art benchmarks from the
PARSEC~\cite{bienia11benchmarking} suite: Blackscholes and Ferret. 
The experiments were executed 5 times in order to remove
discrepancies in the execution time. We used as parallel platform a
machine with four AMD Six-Core Opteron\texttrademark 8425 HE (2100
MHz) chips (24 cores) and 64 GB of DDR-2 667MHz (16x4GB) RAM, running
GNU/Linux (kernel 2.6.31.5-127 64 bits). The machine was running in
multi-user mode, but no other users were in the machine.

\begin{figure}[h!ptb]
\begin{center}
\includegraphics[width=\linewidth]{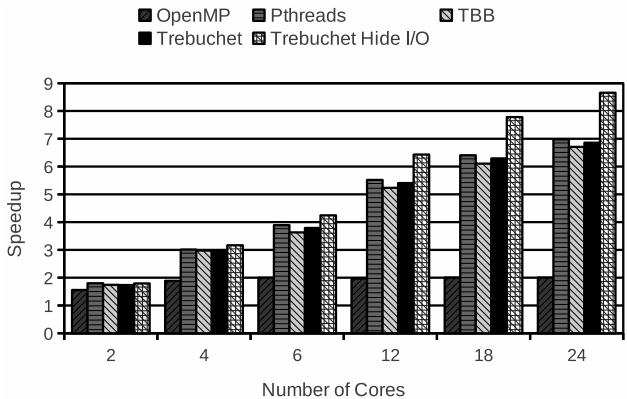}  
\caption{\label{blackscholes}Blackscholes results.}
\end{center}
\end{figure}

We started our study with a regular application: Blackscholes. 
It calculates the prices for a portfolio of European options
analytically with the Black-Scholes partial differential equation
(PDE). There is no closed-form expression for the Black-Scholes
equation, and as such it must be computed numerically. The application
reads a file containing the portfolio. Black-Scholes partial
differential equation for each option in the portfolio can be
calculated independently. The application is parallelized with
multiple instances of the processing thread that will be responsible
for a group of options. Results are then written sequentially to an
output file. The PARSEC suite already comes with 3 parallel versions of 
the Blackscholes benchmark: OpenMP, Pthreads and TBB. We have produced a \emph{Trebuchet}
version of Blacksholes, following the same patterns present in the PARSEC versions to exploit
parallelism. 
However, we observed that we can hide I/O latency and
increase memory locality if we have multiple instances of the input and
output threads. Thus, we have also implemented Blackscholes according to
the example shown at Section \ref{examples}, Figure \ref{hideio}.

Figure \ref{blackscholes} shows the results obtained for the
Blackscholes benchmark. Using TALM language, it is possible to 
obtain good performance (comparable to Pthreads implementations) 
in a simple fashion. However, the flexibility of the language enables 
the programmer to achieve even greater results employing more complex 
techniques of parallelization.

The second benchmark we considered is an irregular application called \emph{Ferret}. This
application is based on the Ferret toolkit which is used for
content-based similarity search. It was developed at Princeton
University, and represents emerging next-generation search engines for
non-text document data types. Ferret is parallelized using the
pipeline model and only a Pthreads version is provided with
PARSEC. However, we had access to a TBB version of Ferret
\cite{Navarro:2009:LBU:1542275.1542358} which is also used in this
experiment.  

First, we have observed that the task size in
Ferret is quite small, and would result in high interpretation
overheads by the virtual machine, specially when using a large number
of cores, where the communication costs become more apparent. 
Therefore, we have adapted the application to 
process blocks of five images per task, instead of one.

\begin{figure}[h!ptb]
\begin{center}
\includegraphics[width=\linewidth]{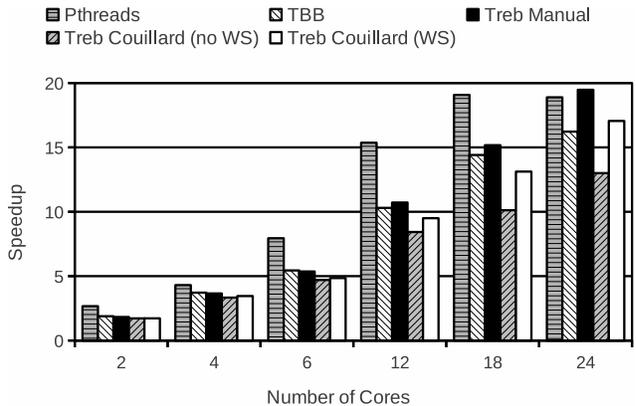}  
\caption{\label{ferret}Ferret results.}
\end{center}
\end{figure}

Our parallel version of ferret uses a pipeline pattern where the 
I/O stages are single super-instructions and processing stages are parallel.
We relied at our work-stealing mechanism (described in Section \ref{talm}) to
perform dynamic load balancing. Results presented in Fig.~\ref{ferret} show 
that our implementation with work stealing (\emph{Treb Couillard (WS)} at the graphic) 
obtains close to linear speedups, for up to 24 cores, and in 
fact performs better than the TBB version, and very close to the speedups 
achieved by the Pthreads version. Also one can note that work stealing added a significant
contribution to the application performance (speedups for \emph{Treb Couillard (no WS)} are
lower).

Moreover, we have also prepared a manually fine-tuned version of ferret,
using over-subscription to rely on the operating system to perform load balancing.
We run \emph{Trebuchet} with 3 times more PEs than the number of used cores and adjust \emph{Trebuchet}'s 
scheduling affinity mechanism to use only the cores necessary for each scenario.
Results show that it is possible to overcome Pthreads' performance. Nevertheless, this minor 
performance gap between a high-level and a manual TALM implementation could be reduced with improvements on 
the work stealing mechanism and addition of code optimization features on \emph{Couillard}.

\Section{Related Work}\label{related}

Data-flow is an long standing idea in the parallel computing
community, with a vast amount of work on both pure and hybrid
architectures~\cite{Swanson2003,10.1109/12.947003,1008761}. Data-flow techniques are widely used in areas
such as internal computer design and stream processing. Swanson's
WaveScalar Architecture~\cite{Swanson2003} was an important influence
in our work, as it was a Data-flow architecture but also showed that
it is possible to respect sequential semantics in the data-flow model,
and therefore run programs written in imperative languages, such as
\texttt{C} and \texttt{C++}. The key idea in WaveScalar
is to decouple the execution model from the memory interface, so that
the memory requests are issued according to the program order. To do
so, WaveScalar relied on compiler to process memory access
instructions to guarantee the program semantics. However, the
WaveScalar approach requires a full data-flow hardware, that has not
been achieved in practice. 

%\emph{Trebuchet} can be thus considered a hybrid Data-flow machine, since it
%allows the compilation of blocks that will be executed in a Von
%Neumann machine, but will be fired according to data-flow. BMDFM
%\cite{PochayevetsThesis} is another hybrid virtual Data-flow machine
%similar to \emph{Trebuchet}. It also supports description of the program in
%both fine and coarse granularities. Fine grained code is written in a functional language based on
%LISP. Regular instructions can be interchanged with
%user-defined instructions written in C, which in BMDFM's terminology
%are called coarse-grained instructions, just like \emph{Trebuchet}'s
%super-instructions. The main difference between both architectures is
%that \emph{Couillard} allows the user to fully describe application in imperative
%language. BMBFM uses various daemons corresponding to the different
%components of the dynamic scheduler, instead of having a virtual machine 
%with PEs implemented as threads of the host machine.

%Program Demultiplexing \cite{1136512} is a different execution
%paradigm where methods or functions are demultiplexed to be executed
%concurrently with the rest of the program, according to data-flow. As
%soon as the demultiplexed method's input parameters are ready, the
%method is fired speculatively. When the control-flow reaches the
%original \emph{call site} for that method, the write operations are
%committed, in case no conflicts are found. Otherwise, the method is
%re-executed.

Threading Building Blocks (TBB) \cite{tbb} is a C++ library designed to provide an abstract layer
to help programmers develop multi-threaded code. TBB enables the programmer to specify
parallel tasks, which leads to a more high-level programming than implementing directly
the code for threads. Another feature of TBB is the use of templates to instantiate
mechanisms such as pipelines. The templates, however, have limitations. For instance,
only \emph{linear} pipelines can be described using the pipeline template.

Another project that relies on code augmentation for parallelization is DDMCPP
\cite{Trancoso2007}. DDMCPP is a preprocessor for the Data Driven Multithreading model
\cite{Kyriacou2006}, which, like TALM, is based on dynamic data-flow.

HMPP \cite{hmpp} is ``an Heterogeneous Multi-core Parallel Programming environment that allows the 
integration of heterogeneous hardware accelerators in a seamless intrusive manner while preserving
the legacy code''. It provides a run time environment, a set of compilation directives and a preprocessor,
so that the programmer can specify portions of accelerator codes, called codelets, that can run at GPGPU, FPGAs, 
a remote machine (using MPI) or the CPU itself. Codelets are pure functions, without side-effects. Multiple codelets implemented for different hardware can exist
and the runtime environment will chose which codelet will run, according to hardware availability and compile directives
previously specified. The runtime environment will also be responsible for the data transfers to/from the hardware components involved
in the computation. 

The Galois System \cite{Kulkarni:2007:OPR:1250734.1250759, Kulkarni:2008:OPB:1346281.1346311, Kulkarni:2008:SSO:1378533.1378575} is an ``object-based
optimistic parallelization system for irregular applications''. It comprises:
\emph{(i)} syntactic constructs for packing optimistic parallelism as iteration
over ordered and unordered sets, \emph{(ii)} a runtime system to detect unsafe
accesses to shared memory and perform the necessary recovery operations and
\emph{(iii)} assertions about methods in class libraries. Instead of tracking
memory addresses accessed by optimistic code, Galois tracks high-level semantics
violation on abstract data types. For each method that will perform accesses to
shared memory, the programmer needs to describe which methods can be commuted
without conflicts (and under which circumstances). Gallois also introduces an alternative method to the commutative checks,
since it may be costly \cite{Kulkarni:2008:OPB:1346281.1346311}.
Shared data is partitioned, attributed do the different processing cores and the system monitors if partitions
are being ``touched'' by concurrent threads (which would raise a conflict). Despite the detection method used, the
programmer needs to describe, for each method that access shared objects, an inverse method that will be executed in case of rollback. 
The runtime system is in charge of detecting conflicts, calling inverse methods and commanding re-execution. 

\Section{Conclusions and Future Work}

We have presented the \emph{Couillard} compiler, that compiles an extension of
the \texttt{C}-language into TALM code. Initial evaluation on
state-of-the-art parallel applications showed TALM code, generated by \emph{Couillard}
and running on Trebuchet (a TALM implementation for multicores),
to be competitive with handcrafted Pthreads and TBB code, up to 24
processors. Evaluation also shows that we can significantly improve
performance by simply experimenting with the connectivity and grain of
the building-blocks, supporting our claim that \emph{Couillard} provides a
flexible and scalable framework for parallel computing.

Work on improving \emph{Trebuchet} continues. Flexible scheduling is an
important requirement in irregular applications, we thus have been
working on improving the work stealing mechanism for \emph{Trebuchet}
runtime environment. Moreover, placement has a strong impact on applications
performance and scalability. We are therefore studying efficient ways
to perform automatic placement on \emph{Trebuchet}.

We are also working on refining \emph{Couillard} and on introducing new
features to the support library. Extending \emph{Couillard} to allow the use of templates
to describe application that fit well known parallel patterns and to enable the use of
\emph{Trebuchet}'s memory speculation mechanisms \cite{trebwammca} are subject of ongoing research. 
This work is based in our experience with porting actual applications to the framework. Thus, 
finding applications that are interesting candidates to be parallelized with \emph{Couillard} is constantly
within our research goals. 

TALM's super-instructions could also be implemented to different hardware, using different languages, as in HMPP \cite{hmpp}, 
as long as there is a way to call them from our virtual machine. Currently, super-instructions are compiled as 
functions in a dynamically linked library, but a interface to call GPGPU or FPGA accelerators and perform data-transfers could also be created in our environment. 
This is subject to on-going work.

\Section*{Acknowledgements}
To CAPES and Euro-Brazilian Windows consortium for the financial support given to the authors of this work.

\bibliographystyle{wscad2009}
\bibliography{talm}

\end{document}